\documentstyle[twocolumn,aps]{revtex}
\input{epsf.sty}

\begin{document}
\draft
\preprint{\it{Submitted to Appl. Phys. Lett.}}
\title{Storage capabilities of a 4-junction single electron
trap with an on-chip resistor}
\author{S.~V.~Lotkhov, H.~Zangerle, A.~B.~Zorin and  J.~Niemeyer\\}
\address{Physikalisch-Technische Bundesanstalt, Bundesallee 100, 
D-38116 Braunschweig, Germany\\}
\date{March 31, 1999}
\maketitle

\begin{abstract}
We report on the operation of a single electron trap comprising a
chain of four Al/AlO$_x$/Al tunnel junctions attached, at one side,
to a memory island and, at the other side, to a miniature on-chip
Cr resistor ($R \approx $ 50 k$\Omega$) which served to suppress
cotunneling. At appropriate voltage bias the bi-stable states of
the trap, with the charges differing by the elementary charge $e$,
were realized. At low temperature, spontaneous switching between
these states was found to be infrequent. For instance, at $T=$70 mK
the system was capable of holding an electron for more than 2
hours, this time being limited by the time of the measurement.
\end{abstract}


The quality and complexity of Single Electron Tunneling (SET)
devices, i.e. the circuits in which the tunneling of electrons is
governed by the Coulomb blockade effect (see, for example, reviews
\cite{Aver} and \cite{Grab}), are steadily growing. The increase of
the number of junctions in these circuits is often motivated by the
necessity to increase the Coulomb energy of the system and to
reduce spontaneous cotunneling events. \cite{AverOdNaz} These
events, that are the higher-order processes, are associated with
electron tunneling occurring in several junctions simultaneously
and quantum-coherently. They set a limit to the accuracy of SET
devices such as turnstiles and pumps (in which an ac signal of
frequency $f$ applied to the gates clocks the transfer of single
electrons), and the storing times of traps (capable to hold an
electron on a memory node of the circuit for long periods). Because
of the cotunneling, a 4-junction SET turnstile and a 3-junction
pump \cite{Urb} realize relation $I = ef$ with an accuracy of about
1\% only, which is insufficient for their applications in metrology
\cite{Flens}. The 4-junction SET traps made by Fulton $et$ $al.$
\cite{Fult} and Lafarge $et$ $al.$ \cite{Laf2} had maximum trapping
times as short as $\sim 1$ s.

In order to make these devices more accurate and reliable (that is
to say, to reduce the probability of one of the most serious source
of errors, namely the cotunneling), the number of series-connected
junctions $N$ should be increased. As was recently demonstrated by
Keller $et \ al.$ \cite{Kel1}, the 7-junction SET pump driven by a
5 MHz signal had an error rate as low as 15 parts in $10^9$. In the
static case, $I = 0$, the duration for an electron to be kept in a
trap also appreciably rises with $N$. For instance, the 7-junction
trap of Dresselhaus $et$ $al.$ \cite{Dress} allowed electrons to be
stored for several hours, this period beeing limited by the
observation time. A similarly high storage capability was found for
the 9-junction trap by Krupenin $et$ $al.$ \cite{KrupLotPr}

On the other hand, increasing of $N$ is not the only way to
suppress the cotunneling. As was theoretically shown by Odintsov
$et$ $al.$ \cite{OBS} for the SET transistor and by Golubev and
Zaikin \cite{GolZai} for the $N$-junction ($N \geq 2$) chain, a
dissipative environment $ |Z(\omega)| = R \gg R_k = h/e^2 \approx
26~$k$\Omega$, can do a good job of suppressing the cotunneling.
(The mechanism of this suppression is qualitatively similar to that
of the Coulomb blockade in a single tunnel junction arising due to
a high serial resistance. This resistor hampers charge relaxation
in the electric circuit and, hence, drastically influences the
tunneling rates.) They found the cotunneling contribution to the
$I-V$ curve of the chain at $T = 0$ and at a small voltage $V$ to
be $$ I \propto V^{2(N+z)-1}, \eqno (1) $$ where $z = R/R_k$. As
can be seen from Eq.(1), parameter $z$ can be regarded as a number
of imagined tunnel junctions $\Delta N$ attached to the
$N$-junction chain and ensuring similar suppression of cotunneling
as resistance $R$.

In this work we pioneered a dramatic reduction of the cotunneling
in a multi-junction circuit (SET trap) by using a dissipative
environment. We utilized an on-chip resistor of about 50 k$\Omega$
(i.e. $z \approx 2$) to reduce cotunneling in a 4-junction chain
with quite ordinary parameters. Since this resistor was roughly
equivalent to two tunnel junctions, we expected the storage
capability of this R-trap to be comparable to that of a 6-junction
trap without resistor.

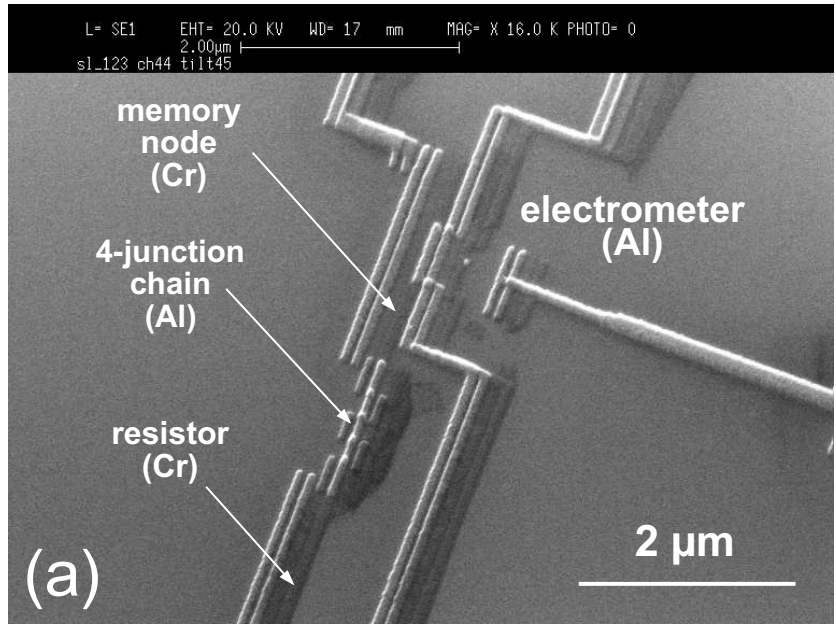
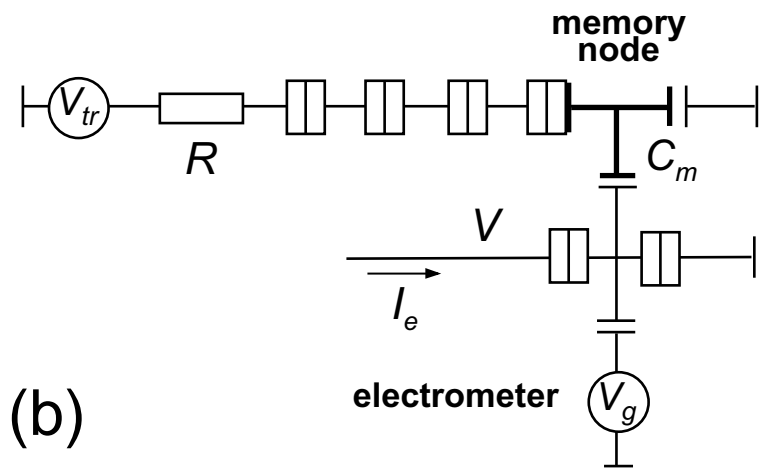
\begin{figure}
\caption{(a) SEM image of the 4-junction R-trap
and measuring SET electrometer. The memory island
(made of Cr) and the Cr resistor have little
contrast in this picture, but their shapes are
similar to those of the Al shadows obtained
with the same mask. (b) The circuit diagram of R-trap. Both,
recharging of the memory island and varying of the offset charge
on the electrometer island could be done by the voltage
applied either to the gate, $V_g$, or to the resistor, $V_{tr}$.}
\label{layout}
\end{figure}

The sample, comprising the trapping array itself and the readout
SET electrometer positioned near the memory island (see
Fig.~\ref{layout}), was fabricated by the well-established shadow
deposition technique \cite{NiemDol} through the trilayer mask with
"hanging bridges" patterned by e-beam lithography and reactive-ion
etching. Through the same mask, at three different angles
($-23^{\rm{o}}, 0^{\rm{o}}$ and $+12^{\rm{o}}$), we consequently
deposited $in$ $situ$ three metal layers: Cr (8 nm thick), then Al
(30 nm), and after oxidation again Al (35 nm). All tunnel junctions
on the chip had the same nominal dimensions, 80 nm$ \times$ 80 nm,
and, as was found from measurements of the electrometer transistor,
their intrinsic capacitance was $C \sim 160$ aF and their tunnel
resistance $R_t \sim 70$ k$\Omega$. The memory island of the trap
was nominally 80 nm$ \times$ 2.2 $\mu$m and had a self-capacitance
$C_m$ in the range of 100 aF. Each of three inner islands was about
80 nm$ \times$ 300 nm in size. The Cr resistor was 8 $\mu$m long,
80 nm wide and its resistance ($R \approx 50$ k$\Omega$) was
evaluated from the measurement of similar single resistors
fabricated apart on the same chip. \cite{Comm}

The measurements were carried out in a dilution refrigerator within
the temperature range $T = 70-170$~mK. A magnetic field of 1 T was
applied perpendicular to the chip in order to keep Al parts of the
circuit in the normal state. In the biasing lines we used
$\pi$-filters against rf noise and the Thermocoax filters against
the microwave frequency noise. The latter filters were the pieces
(1 m long) of the coaxial cable lines (of diameter 0.5 mm) having
considerable losses ($>$ 100 dB at $f > 10$ GHz). \cite{Zor} The
cables were thermally anchored at the mixing chamber temperature
and fed through into a shielding case containing the sample holder.
In order to characterize the quality of filtering by an effective
temperature $T_e$ referred to the sample, we had measured a test
sample comprising a SET electrometer coupled to an electron box
\cite{Laf1}. The value $T_e = 50-60$ mK had been found from the
switching characteristics of the box at $T=20$~mK.~\cite{ZanTh}

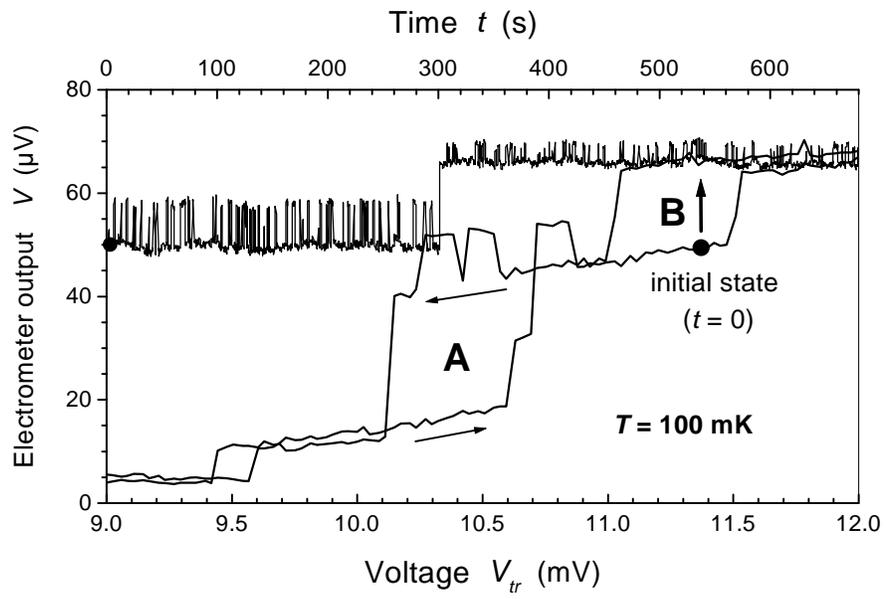
\begin{figure}
\caption{Voltage $V$ across the electrometer
(fed by a small current $I_e = 2$ pA in order not to disturb
the trap) against voltage $V_{tr}$.
The sweep (the directions
are shown by thin-line arrows) was made at the rate of 0.15 mV/s.
The fluctuations of the
background charge (with the most prominent two-level fluctuator
causing sporadic jumps in the offset charge of the electrometer
$\delta Q \approx 3.5\times 10^{-2}e$) are clearly
seen in the $V-V_{tr}$ plot
as well as in the time-trace (presented in the upper part
of the figure).
Spontaneous switching within loop B (shown
by the heavy-line arrow), which occurred after a lapse of 5 minutes
is clearly seen in the time trace.
Note that the temperature in this experiment was
elevated up 100 mK and the bias point was moved out of
the loop center to demonstrate the effect of switching
on a suitable time scale.}
\label{loops}
\end{figure}

The presence of the memory island near the electrometer clearly
manifested itself as regular jumps of output voltage $V$ in both
the $V-V_g$ and the $V-V_{tr}$ characteristics of the electrometer.
The change of the polarization charge $\delta Q$ on the
electrometer island caused by charging of the memory island by an
elementary charge was found to be $(7.6 \pm 0.5)\times 10^{-2} e$.
This value was large enough to reliably monitor integer jumps of
the charge on the memory node against a background noise. The
modulation curves recorded for ramping $V_g$ or $V_{tr}$ in
positive and negative directions formed typical hysteresis loops
(the memory effect) \cite{Fult,Laf2,Dress,KrupLotPr}, indicating
that an electron enters into, or leaves, the memory island at the
different values of bias. Neighboring loops were apart from each
other and their positions were rather stable in time (i.e. the
drift of the background charge was reasonably small). This proved
that our trap had exactly two charge states within each loop (see
Fig.~\ref{loops}). Since the inner islands of the chain were not
supplied with individual tuning gates (as it was done in, e.g.,
Ref. \cite{Laf2}), we were not able to maximize the width of the
loops by applying appropriate voltages which could compensate the
random offset charges and, thereby, increase the energy barrier
$\Delta U$ separating the states. Instead, we simply chose the
loops with larger widths, which presumably corresponded to
favorable distributions of the offset charges ensuring larger
$\Delta U$.

The holding time within some of the loops (for instance, those
marked in Fig.~\ref{loops} as loops A and B) was found to be
actually long. At $T = 70$ mK, the hold time of the "upper" and
"bottom" states for the bias corresponding to the centers of these
loops was determined by the time of observation (more than 2 h). In
another measurement, we ramped voltage $V_{tr}$ over loop B,
starting from 10.8 mV and rising up to 11.8 mV at an average rate
of $5\times 10^{-8}$ V/s (i.e. 1 mV per 5.5 h). In response to the
change of $V_{tr}$ the system switched from the initial "bottom"
state to the "upper" state only once, i.e. the induced transition
took place. (It occurred after a lapse of 2.5 h at $V_{tr} \approx
11.25$~mV, i.e. at the value inside the loop recorded at the normal
sweep rate of $1.5 \times 10^{-4}$ V/s.) After that the system
remained in that state until the end of the ramp, i.e. for about 3
h. At $T = 100$~mK, transitions occurred on a reasonable time scale
of about $10^3$ s when $V_{tr}$ was adjusted to the centers of the
loops. When the bias was changed such that the energy of the
occupied state exceeded the energy of the unoccupied state, induced
switching occurred faster. Such case is illustrated by the time
trace in Fig.~\ref{loops}.

In order to evaluate the barrier height we elevated the temperature
and thereby made thermally activated switching between the
bi-stable states more intensive. We then found from the Arrhenius
plot the activation energy $\Delta U = 240
\pm 20~\mu$eV, what corresponds to $\approx 2.8 \pm 0.2$ K, which
characterizes the pair of the charge states for loop B. This value
is plausible for the evaluated parameters of the trap if we assume
non-zero offset charges on the inner islands on the chain. The rate
$\Gamma_{th}$ of thermally activated transitions at $T \leq 100$ mK
for such barrier, extrapolated from higher temperature
measurements, was found to be below $2 \times 10^{-6}~\rm{s}^{-1}$.

For a crude evaluation of the cotunneling rate $\Gamma_{cot}$ we
used formula (9b) obtained by Golubev and Zaikin \cite{GolZai},
assuming the net energy change to be $\Delta E \sim k_BT$. For $T =
70-100$ mK we obtained $\Gamma_{cot} \sim
10^{-11}-10^{-9}~\rm{s}^{-1}$ for our sample .
The maximum rate of leakage evaluated from our short-term
measurements was $\Gamma \sim 10^{-4}~\rm{s}^{-1}$, i.e. it is by
several orders of magnitude higher than $\Gamma_{cot}$. (Note that
the experimental values of $\Gamma$ are always much larger than
theoretical estimations of both the thermal activation and
cotunneling. \cite{Fult,Laf2,Kel1,Dress,KrupLotPr}) On the other
hand, the value obtained for $\Gamma$ was considerably lower than
that evaluated \cite{GolZai} for the case of a 4-junction trap with
similar parameters but without resistor, $viz.$ $\Gamma'_{cot} \sim
10^{-3}-10^{-2}~\rm{s}^{-1}$. This fact points to the crucial role
the resistor plays in the improvement of the trap's storage
capability. In particular, the characteristics of this R-trap are
comparable to those of the well-characterized 7-junction SET pump
\cite{Kel1,Kel2Kautz} in the hold mode of operation. That device
had tunnel junctions of $C \approx 220$ aF and $R_t \approx 470$
k$\Omega$, and 6 tuning gates allowed the leakage to be reduced
down to $(0.3-20)\times 10^{-4}~\rm{s}^{-1}$ at $T=40$ mK.

In summary, we have fabricated and characterized a 4-junction
electron trap with an on-chip resistor. We have demonstrated the
device is capable of storing a fixed number of electrons on its
memory node on the appropriate time scale. We believe that almost
similar storing capability can be achieved for a trap consisting of
only 2 tunnel junctions and equipped with a resistor of $R \approx
100$ k$\Omega$ which yields $z \approx 4$. (Our estimation gives in
that case $\Gamma_{cot} \sim 10^{-7}~\rm{s}^{-1}$.) In such
2-junction trap, the electrostatic barrier (and thereby the storing
time) could be effectively controlled by a gate polarizing the
island between the junctions. The obtained result is extremely
encouraging with respect to constructing a fewer-junction SET
pump/turnstile device with resistors, which is our next goal. For
instance, a 3-junction pump supplied with two 50 k$\Omega$
resistors, yielding $\Delta N = z = 2 \times R/R_k \approx 4$,
could be as accurate as its 7-junction counterpart without
resistors. The obvious advantage of such an R-pump is the minimum
number of gates (two) and, hence, much simpler rf-drive (two
harmonic signals with a fixed phase shift). Finally, the total
drift of a working point of the device caused by the fluctuations
of the offset charges on 2 islands should be apparently weaker than
in the case of 6 similar islands.

The authors thank V.~A.~Krupenin for valuable discussions. The work
is supported in part by the EU (MEL ARI Research Project $22953 -$
CHARGE) and the German BMBF (Grant No. 13N7168).

\begin{references}
\bibitem{Comm} These resistors exhibited practically linear $I-V$ curves
at low $T$. The self-capacitance of these strips $C_0$ was found to be
about 550 aF, i.e. $\approx$ 70 aF per 1 $\mu$m of length. From this we have
concluded that our $RC$ line behaved as a pure ohmic resistance for frequencies
$\omega < \Omega_{RC}= (RC_0)^{-1} \approx 4 \times 10^{10}s^{-1}$ or
$V < \hbar \Omega_{RC}/e \approx 30~\mu$eV.
\end{references}
\end{document}